\documentclass[manuscript]{emulateapj}
\usepackage{graphicx}
\usepackage{latexsym}
\usepackage{amssymb}
\usepackage{float} 
\usepackage{threeparttable}

\shorttitle{Color-Magnitude Distribution of Face-on Nearby Galaxies in SDSS DR7}
\shortauthors{Jin et al.}

\begin{document}


\title{Color-Magnitude Distribution of Face-on Nearby Galaxies in SDSS DR7}



\author{Shuo-Wen Jin}
\affil{Purple Mountain Observatory, Chinese Academy of Science, 210008, China}

\author{Qiusheng Gu}
\affil{School of Astronomy and Space Science, Nanjing University, Nanjing, 210093, China; \\qsgu@nju.edu.cn}

\author{Song Huang}
\affil{School of Astronomy and Space Science, Nanjing University, Nanjing, 210093, China}

\author{Yong Shi}
\affil{School of Astronomy and Space Science, Nanjing University, Nanjing, 210093, China}

\author{Long-Long Feng}
\affil{Purple Mountain Observatory, Chinese Academy of Science, 210008, China}

\begin{abstract}

We have analyzed the distributions in the color-magnitude diagram (CMD) of a large  sample of  face-on galaxies  to  minimize the  effect of  dust extinctions on galaxy color.  About 300 thousand galaxies  with $log(a/b) < $  0.2 and redshift $z < 0.2$ are selected from the  SDSS DR7 catalog.  Two methods  are employed to investigate the distributions of  galaxies  in  the  CMD  including  1-D  Gaussian  fitting  to  the distributions in  individual magnitude  bins and 2-D  Gaussian mixture model (GMM)  fitting to galaxies as a  whole. We find  that in the 1-D  fitting only two Gaussians  are not  enough to  fit galaxies  with the excess present between the  blue cloud and  the red  sequence. The fitting to  this excess defines the  centre of the  green-valley in the local universe  to be  $(u-r)_{0.1} = -0.121M_{r,0.1}-0.061$. The  fraction of blue  cloud and red  sequence galaxies  turns over  around $M_{r,0.1} \sim -20.1$ mag, corresponding  to stellar  mass of $3\times10^{10}M_\odot$. For the  2-D GMM fitting,  a total of  four Gaussians  are required,  one for  the blue cloud, one  for the red sequence  and the additional two  for the green valley.  The fact  that  two  Gaussians are  needed  to describe  the distributions of galaxies  in the green valley is  consistent with some models that argue for two different evolutionary paths  from the blue cloud to the red sequence. 

\end{abstract}


\section{Introduction}

The distribution of galaxies in the color-magnitude diagram (CMD) provides a powerful tool to investigate the evolution of galaxy populations.
A remarkable feature of the CMD is a robust bimodality, which divides the galaxy population into a ``blue cloud'' (or blue sequence) and a ``red sequence''. The bimodality is seen in the optical colors \citep{Strateva_2001, Blanton_2003c}, UV$-$optical colors \citep{Wyder_2007}, the 4000 \r{A} break ($D_n4000$; \citealt{Kauffmann_2003a}), and spectral type \citep{Madgwick_2002}. Galaxies in ``red sequence" are quiescent, bulge-dominated galaxies \citep{Blanton_2009}, while the ``blue cloud" is characterized by star-forming, disk-dominated galaxies. 
Between these two sequences, there is a region called the ``green valley".

\citet{Baldry_2004} explored the distribution of galaxies in the $(u-r)$ versus $M_r$ diagram for low-redshift SDSS samples. Their galaxies separate into ``blue cloud" and ``red sequence", and the distribution of $(u-r)$ color at each absolute magnitude bin is well fitted by the sum of two Gaussians. However, \citet{Wyder_2007} showed that the $(NUV-r)$ color distribution  at each $M_{r}$ can not be fitted well by the sum of two Gaussians due to an excess of galaxies between the blue and red sequences. They utilized Balmer decrements and the Dust-SFH (star formation history)-Color relation \citep{Johnson_2006} to correct the extinction of each galaxy, there still remain galaxies in the green valley region between two sequences. Thus, galaxies in the green valley region may not be a simple mixture of blue and red galaxies.

The understanding of the CMD color bimodality is also complicated by galaxy dust extinction. \citet{Salim_2009} found that many green valley galaxies are simply dust-obscured actively star-forming (SF) galaxies.
However, there still exist 24 $\mu$m detected galaxies, some with LIRG-like luminosities, which have little current SF. They belong to green valley or even the red sequence because of their SF history, not just dust reddening.

The CMD bimodality is already in place at $z\sim1$ \citep{Cooper_2006}, with color becoming bluer at higher redshift \citep{Blanton_2006, Willmer_2006}. Based on the DEEP2 and COMBO-17 surveys, Faber et al. (2007) argued that the number density of blue galaxies is more or less constant from $z\sim1$ to 0, while the number density of red galaxies has increased. This work supports that the red sequence has grown in mass by a factor of 3 since $z\sim1$. A plausible scenario is that the growth of red galaxies was triggered by quenching star formation in blue galaxies, which caused them to migrate into the red sequence \citep{Bell_2004}. In addition, galaxies may also be moving from the lower end of the red sequence to the blue cloud through accreting gas-rich dwarf galaxies \citep{Faber_2007}. 

Studies of galaxy morphologies show that red sequence is dominated by spheroidal galaxies with S{\'e}rsic index n=4, while the blue cloud is occupied by disk-dominated galaxies with S{\'e}rsic index n=1 \citep{Driver_2006}.
\citet{Mendez_2011} investigated the morphologies of green valley galaxies from the AEGIS survey and found that most green valley galaxies are not classified as mergers and that the merger fraction in the green valley is lower than that in the blue cloud. \citet{Lackner_2012} presented a set of bulge-disc decompositions for a sample of 71825 SDSS main-sample galaxies and found that the majority of green valley galaxies are bulge+disc galaxies, and that the integrated galaxy color is driven by the color of galaxy disks.
 
In general, blue galaxies with star formation being quenched will evolve from the blue cloud to the red sequence, passing through the green valley that thus represents an intermediate phase of this quench process.
Different mechanisms have been proposed to cease star formation in blue galaxies, such as mergers \citep{Bell_2004, Hopkins_2010}, AGN feedback \citep{Croton_2006, Martin_2007, Schawinski_2010}, morphological quenching \citep{Martig_2009}, cold flows accretion and shock heating \citep{Dekel_2006, Cattaneo_2006} (see \citealt{Peng_2010} for a reccent review).

\citet{Peng_2010,Peng_2012} investigated the quenched fraction of galaxies as a function of local density, stellar mass, and redshift. 
They parameterized galaxy quenching as fully separable ``environment quenching" and ``mass quenching", which are directly associated with the quenching processes of satellite and central galaxies in group. 
This model is successful in predicting the mass function of passive and star-forming galaxies.
These effects may also be reflected in the color-magnitude distribution of galaxies, we will try to find some evidence of these affects in our analysis.

In this paper, we focus on fitting CMD by the use of different methods. Our study aims at a better understanding of the ``green valley" and may answer the question whether the ``green valley" is dominated by one component. Since different quenching mechanisms will produce different distribution of galaxies in color-magnitude space, the fine structure of CMDs may also provide valuable clues about galaxy quenching mechanisms. Previous work always focus on the whole galaxy population without identification, which may conceal the fine structure of CMDs due to dust extinction, so we selected a nearly face-on galaxy sample to minimize the effect of dust.
 
This paper is organized as follows. Section 2 describes the sample selection in this work. In Section 3, we investigate the color-magnitude distribution and show 1D and 2D Gaussian fitting results. Section 4  discuss the implication of the results. Section 5 is the conclusion.
Throughout this paper, we assume a flat $\Lambda$CDM cosmology with a matter density $\Omega_{m}$=0.3, cosmological constant $\Lambda$=0.7 and Hubble's constant $H_0$=100$kms^{-1}Mpc^{-1}$ (i.e., $h=1$).

\section{The Sample}
The photometric and spectroscopic data used in this paper are taken from the main galaxy sample in Sloan Digital Sky Survey (SDSS) DR7 \citep{York_2000, Strauss_2002, Abazajian_2009}. The sample selection criteria are as follows:

(1) $r$ band apparent magnitude $m_r < 17.77\ mag$ with $ugriz$ flux signal-to-noise ratio $S/N>5$. The criterion of S/N can reduce outliers. By comparing the color distribution of bright galaxies and the overall population, this criterion doesn't generate significant selection effects.

(2) $log(a/b)<0.2$, $a/b$ is the ratio of major axis to minor axis of de Vaucouleurs' profile. Gas and dust tend to reside in the disk of spirals, the colors of spirals are seriously reddened by dust attenuation. For example, the total extinction from face-on to edge-on is about 0.7, 0.6, 0.5 and 0.4 $mag$ for the $ugri$ passbands \citep{Masters_2010b}, so we choose the nearly face-on galaxies to minimize the dust reddening effect.

In total, our sample consists of 329384 galaxies. To ensure a broader span of luminosity and the completeness of color in each magnitude bin, we divide galaxies into 12 bins from $M_{r,0.1}=-18.50\sim-21.5\ mag$ and $0.02<z<0.18$ with bin size of 0.25 $mag$, as shown in Figure 1 ($M_{r,0.1}$ is the absolute magnitude in $r$ band $k-corrected$ to $z=0.1$). 
The galaxy number and median redshift in each bin are shown in Table 1. We will analyse the color distribution of galaxies in each bin.

\begin{figure}
\includegraphics[width=3.2in]{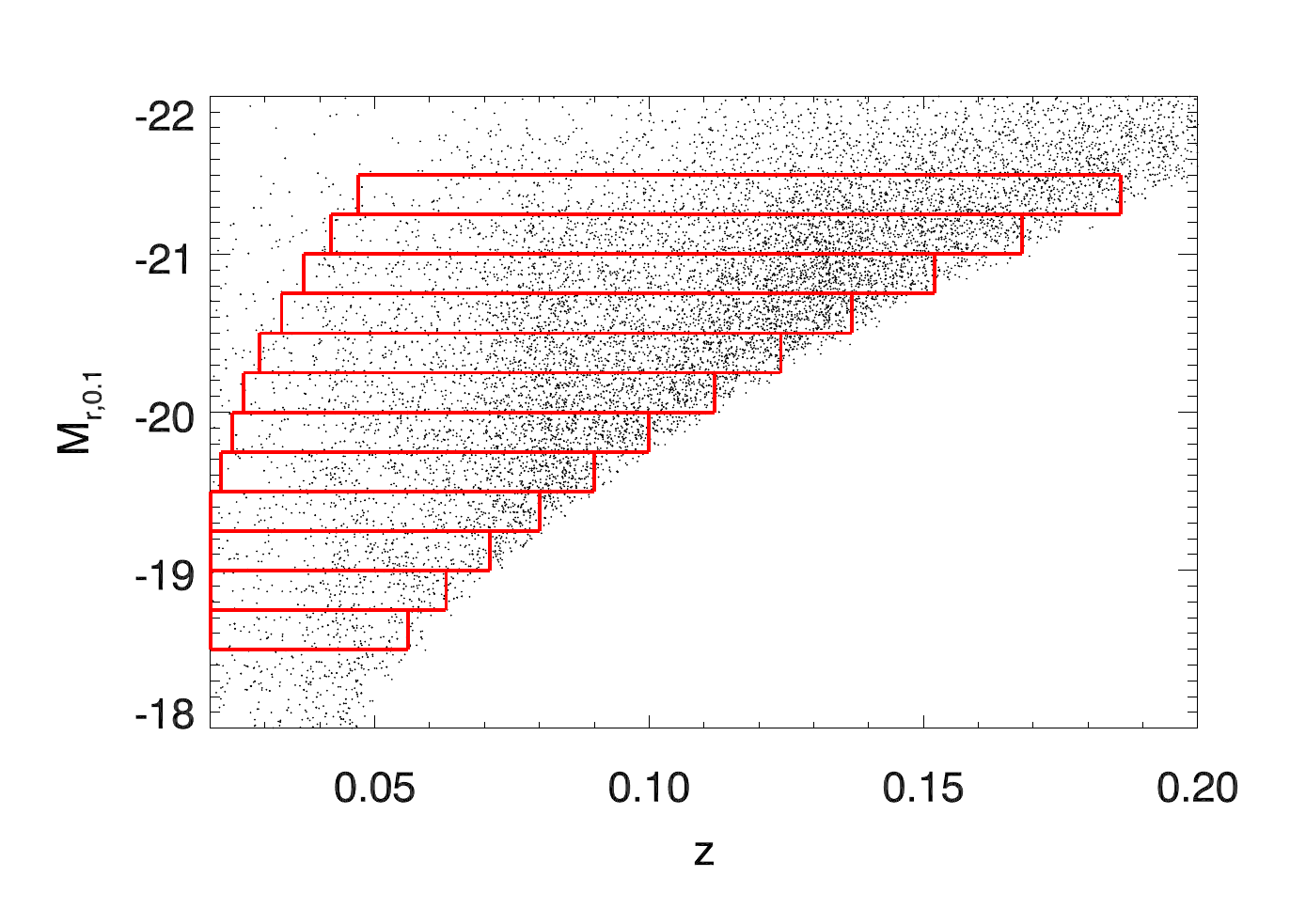}
\caption{We select galaxies in each red box to ensure the completeness of color in each magnitude bin. Black dots show a random subsample of SDSS face-on $[log(a/b)<0.2]$ main galaxy sample. $M_{r,0.1}$ is the absolute magnitude in $r$ band which have been $k-corrected$ to $z=0.1$. The galaxy number and median redshift in each bin are shown in Table.1.}
\label{fig.1}
\end{figure}

\begin{table}
 \centering
 \begin{minipage}{83mm}
  \caption{The galaxies in each red bin of Figure.1}
  \begin{tabular}{cccc@{}}
  \hline
   $M_{r,0.1}$ & $z$ & $Galaxy\ Number$ & $\overline{z}$\\
  \hline

$[-18.50,-18.75]$ & $[0.020,0.056]$ & 3541& 0.042\\
$[-18.75,-19.00]$ & $[0.020,0.063]$ & 4600& 0.047\\
$[-19.00,-19.25]$ & $[0.020,0.071]$ & 6715& 0.053\\
$[-19.25,-19.50]$ & $[0.020,0.080]$ & 9800& 0.061\\
$[-19.50,-19.75]$ & $[0.022,0.090]$ & 13265& 0.068\\
$[-19.75,-20.00]$ & $[0.024,0.100]$ & 16225& 0.074\\
$[-20.00,-20.25]$ & $[0.026,0.112]$ & 20116& 0.083\\
$[-20.25,-20.50]$ & $[0.029,0.124]$ & 23683& 0.092\\
$[-20.50,-20.75]$ & $[0.033,0.137]$ & 27912& 0.103\\
$[-20.75,-21.00]$ & $[0.037,0.152]$ & 30836& 0.114\\
$[-21.00,-21.25]$ & $[0.042,0.168]$ & 31382& 0.126\\
$[-21.25,-21.50]$ & $[0.047,0.186]$ & 29832& 0.139\\

\hline
\end{tabular}
  \begin{tablenotes}
    \footnotesize
    \item[1] $\overline{z}$ is the galaxies' median redshift in each bin.
  \end {tablenotes}
\end{minipage}
\end{table}

Using the New York University Value-Added Catalog (NYU-VAGC; \citealt{Blanton_2005}) and  MPA-JHU catalog, we obtain the physical parameters for our sample, such as $u,g,r,i,z$ absolute magnitudes k-corrected to $z=0.1$ \citep{Blanton_2003a}, stellar mass, and $D_n4000$ \citep{Kauffmann_2003a, Salim_2007}. 
We use {\sc modelMags} for galaxy magnitude and color in this paper, which are derived from the best-fitting exponential or de Vaucouleurs galaxy profile. 
{\sc modelMags} have occasional problems recovering accurate magnitudes for galaxies with mixed morphologies, and {\tt petro} magnitude \citep{Petrosian_1976} provides a better measure for flux \citep{Taylor_2011,Simard_2011}, but the use of {\sc modelMags} do not change any of the results that we present in Section 3.

Although galaxies in each magnitude bin cover certain redshift ranges, the effect of redshift evolution in color should be negligible given the overall redshift range of  the whole sample from 0.0 to 0.2.

\begin{figure*}
\centering
\includegraphics[scale=0.53]{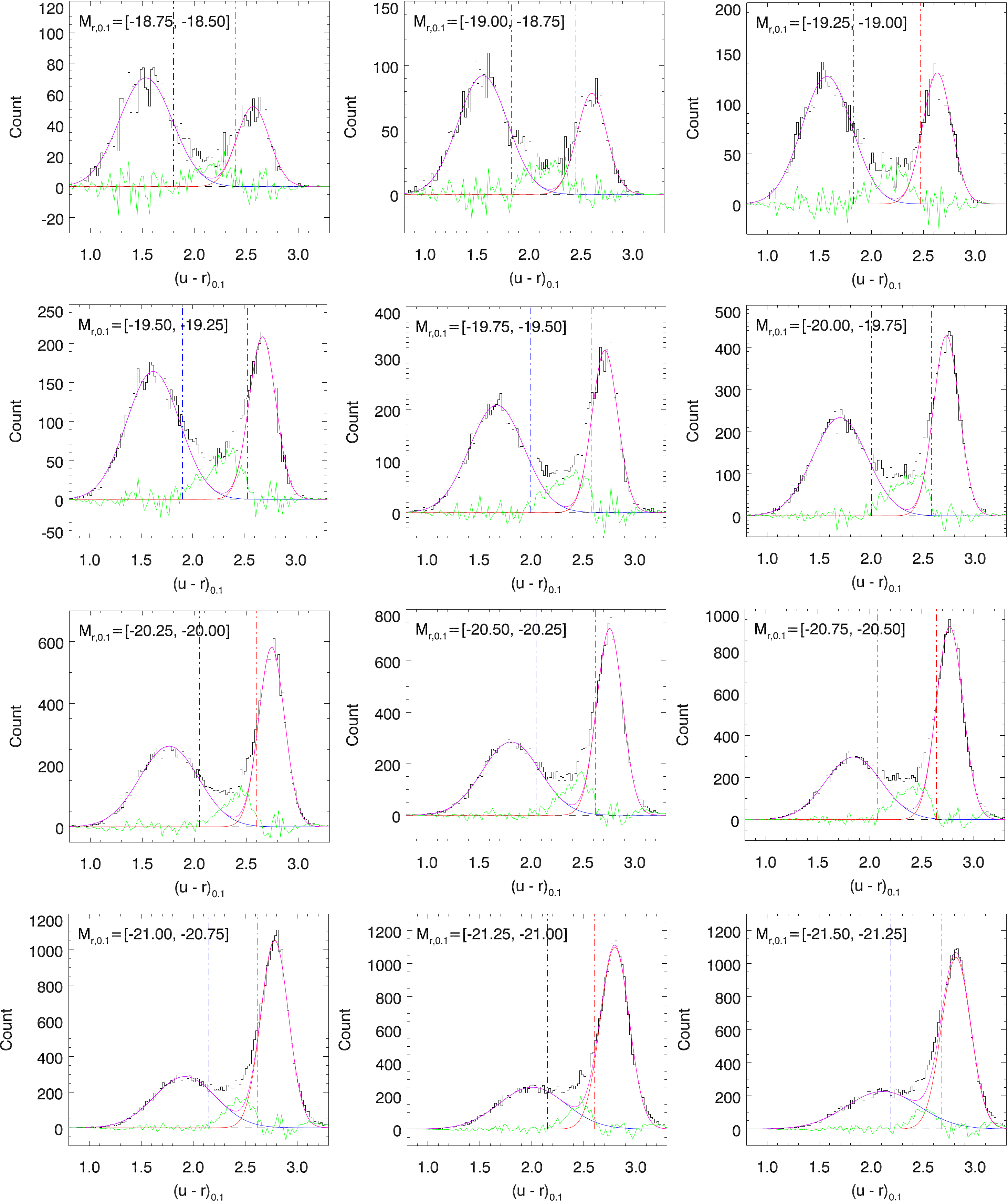}
\caption{The $(u-r)$ color distribution in 0.25 $mag$ absolute magnitude bins. The magnitude range is labeled in the upper left of each figure from -21.50 to -18.50 $mag$. The black histogram is the $(u-r)$ distribution in each magnitude bin, the color histogram's bin is 0.02 $Mag$. The blue and red dash lines mark the color boundaries when fitting Gaussian distribution to blue cloud and red sequence. The solid blue and red lines are the Gaussian fitting results for blue cloud and red sequence, the solid purple line is the sum of two Gaussians. Finally, the solid green marks the residual distribution between actual color distribution and the sum of two Gaussians.}
\label{fig.2}
\end{figure*}

\section{Results}
\subsection{Color Distribution Fitting}

\begin{figure}
\centering
\includegraphics[width=3.0in]{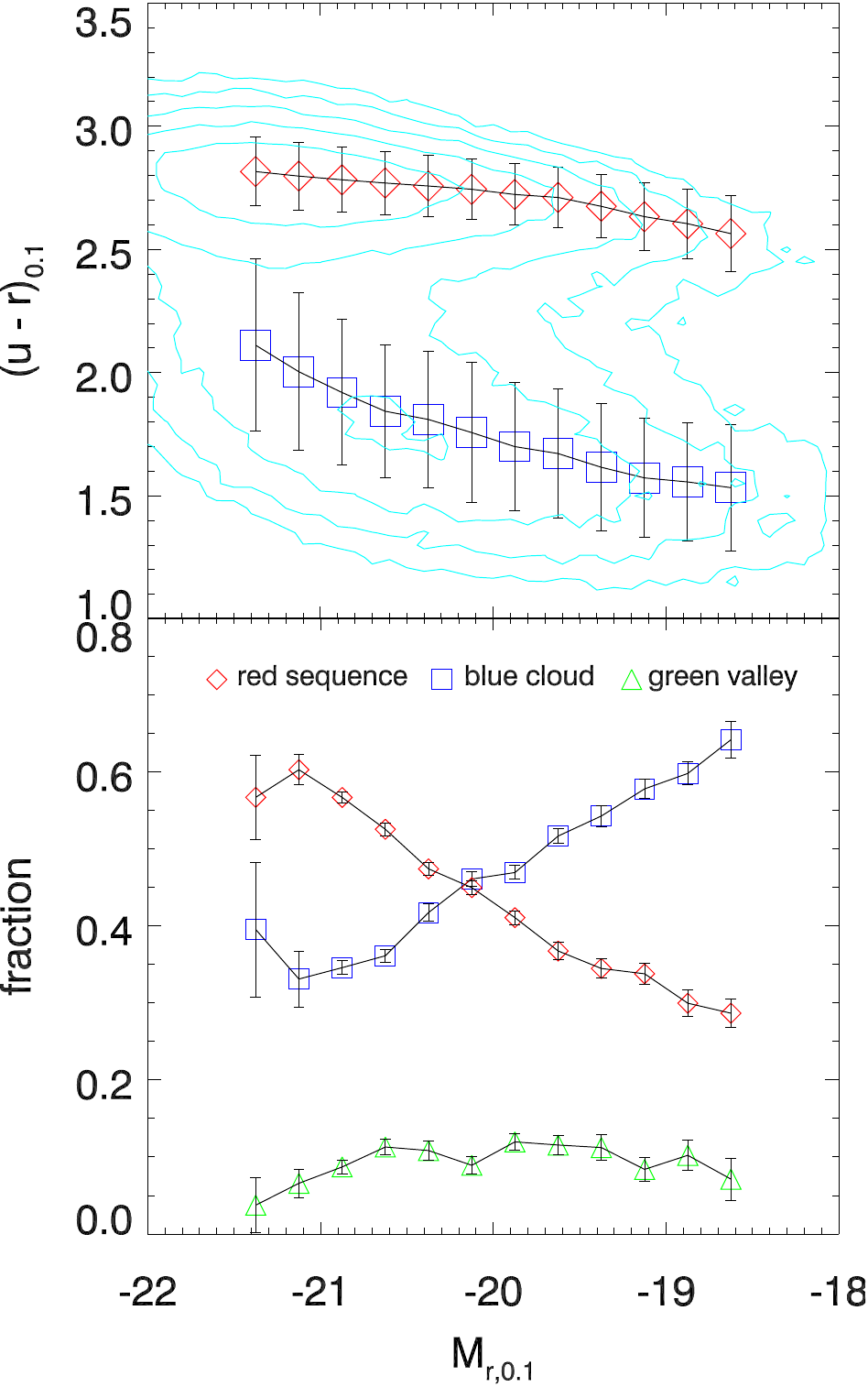}
\caption{The top panel plots the color-magnitude diagram (CMD) of face-on galaxies and the best fit to each sequence in Figure 2. The contour is all face-on galaxies distribution in CMD, the blue boxes mark the median color of blue sequence in each magnitude bin, the red diamonds mark the red sequence. The error bar of each point refers to the 1 $\sigma$ width of the Gaussian. The lower panel shows the fraction of blue cloud galaxies, red sequence galaxies and green-valley galaxies, the error bars are calculated by Bootstrap resampling method.}
\label{fig.3}
\end{figure}

As shown in Figure 2, the distributions of galaxies in each magnitude bin show apparent two peaks except for two brightest bins where the distribution profile is characterized by a red peak plus a blue tail. We fit a single Gaussian to blue and red peaks. Respectively, we select the left 0.15 $mag$ of red peaks as the red boundaries and select the right 0.25 $mag$ of blue peaks as the blue boundaries. For the two brightest bins, since a very small blue fraction, we artificially adjusted the boundaries until these two peaks are well fitted,
the color boundaries are plotted as the dashed blue and red lines in each magnitude bin in Figure 2. 

As shown in Figure 2, two Gaussians accurately fit the blue cloud and red sequence, but the color distribution apparently cannot be fitted well by the sum of 2 Gaussians  in each magnitude bin because of the obvious excess between the two sequences. We define the excess between two sequences as the ``green-valley galaxy" population.

The parameters of the best fit are shown in Figure 3. In the top panel, the contours illustrate the nearly face-on galaxies distribution in CMD, the blue boxes mark the median color of blue sequence and the red diamonds mark the red sequence, the error bar of each point gives the 1 $\sigma$ width of the Gaussian fitting. 

To quantify each sequence, the fraction of blue cloud (red sequence) is defined as the ratio of blue cloud (red sequence) galaxies number to total number in each magnitude bin, and the fraction of green-valley galaxies is defined as the complementary set of them. In the bottom panel of Figure 3, the fraction of red sequence rises with luminosity while the fraction of blue cloud galaxies declines, they turn over around  $M_{r,0.1}\sim-20.1\ mag$. The trend of both red and blue fraction in the brightest bin may be artificial since a very small blue fraction makes two-Gaussian fitting difficult (see last panel of Figure 2). The fraction of green valley galaxies remains almost constant around 10\% except for the two brightest bins.

Interestingly, we find the mean stellar mass of galaxies in $M_{r,0.1}\sim-20.1\ mag$ bin is close to the characteristic mass $10^{10.5}M_{\sun}$ \citep{Kauffmann_2003a} , that is, the fraction of red sequence and blue cloud cross over at the characteristic mass $M^*$.

We also fit the $(u-g)$ color distribution in the same way and find the same results. For the $(g-r)$ color distribution, the blue cloud and red sequence are too close to apply this methodology. We have already applied redshift evolution correction to the data \citep{Blanton_2003b}, which makes no change to the results.

\begin{figure}
\includegraphics[width=3.4in]{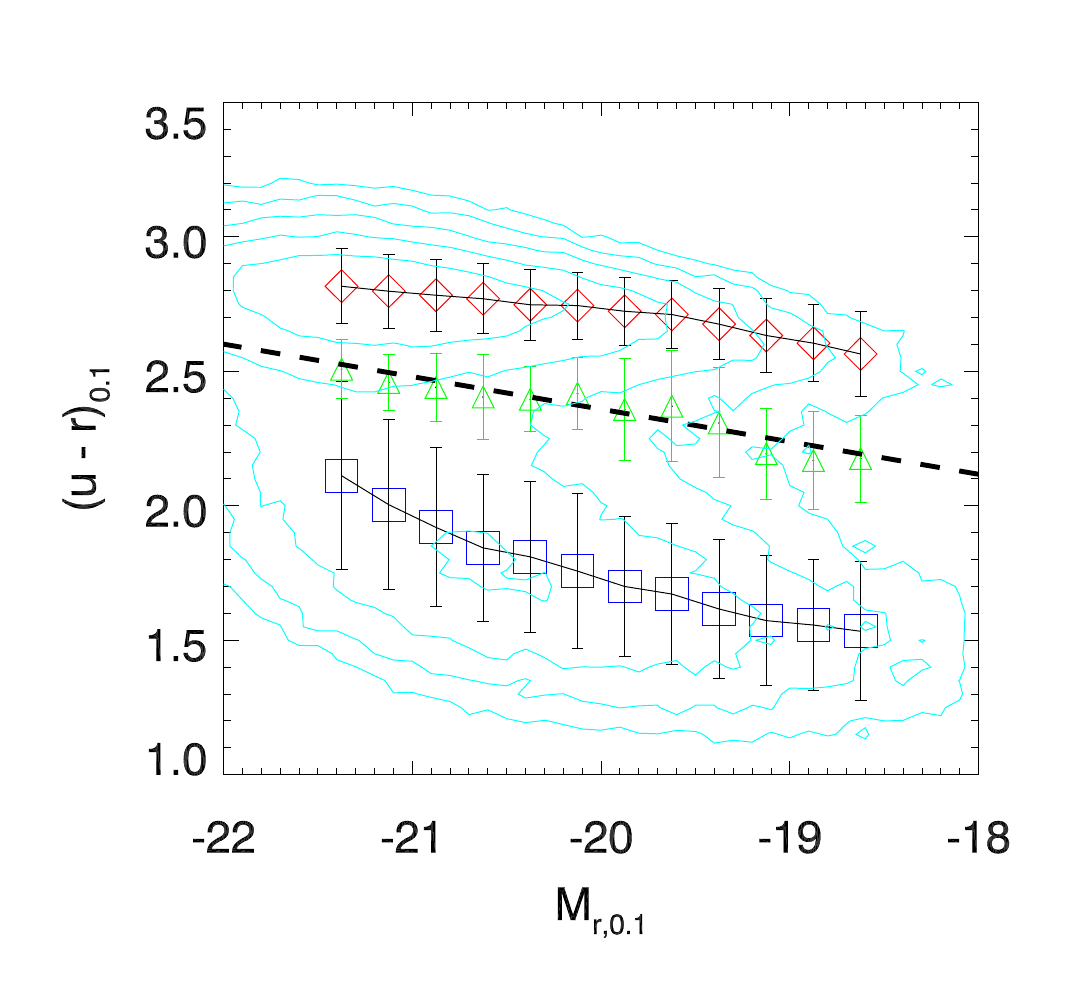}
\caption{The contours, boxes and diamonds are the same as the case in the top panel of Figure 3, the green triangles and their error bars refer to the green valley location in each magnitude bin which is calculated by fitting the green residuals in Figure 2 by a Gaussian, the dash black line is the liner fit of green triangles.}
\label{fig.4}
\end{figure}

In order to derive the location of green-valley galaxies, we can roughly fit the excess (green line in Figure 2) by a Gaussian in each magnitude bin, the median color and 1 $\sigma$ region are plotted as triangles and error bars in Figure 4. We fit the triangles with a line, which plotted as the black dash line in the left panel, that is the expression of the centre of the green valley:

\begin{equation}
\centering
(u-r)_{0.1}=-0.121M_{r,0.1}-0.061.
\end{equation}

\subsection{Color-Magnitude Distribution} 

Above, we have studied the color distribution in each magnitude bin. However, the color-magnitude distribution is actually a two dimensional distribution, and the evolutionary pathway of galaxy may span several magnitude bins. The two dimensional fitting for CMD may provide us more information, so we apply Gaussian mixture model (GMM) to fitting the color-magnitude diagram and color-mass diagram directly. This research made use of astroML, a community-developed core Python package for Astronomy \citep{VanderPlas_2012}.

The discontinuity of the volume limited samples in section 2 makes them inappropriate for GMM fitting, so we select a subsample from our main sample with criteria $0.04<z<0.08$, galactic extinction in the $u$ band less than 0.3 $mag$ and in the $r$ band less than 0.5 $mag$, which are used to reduce outliers. As shown in Figure 5, this subsample contains 67499 galaxies and  is complete to $M_{r,0.1}<-19.2\ mag$. 

\begin{figure}
\centering
\includegraphics[width=3.2in]{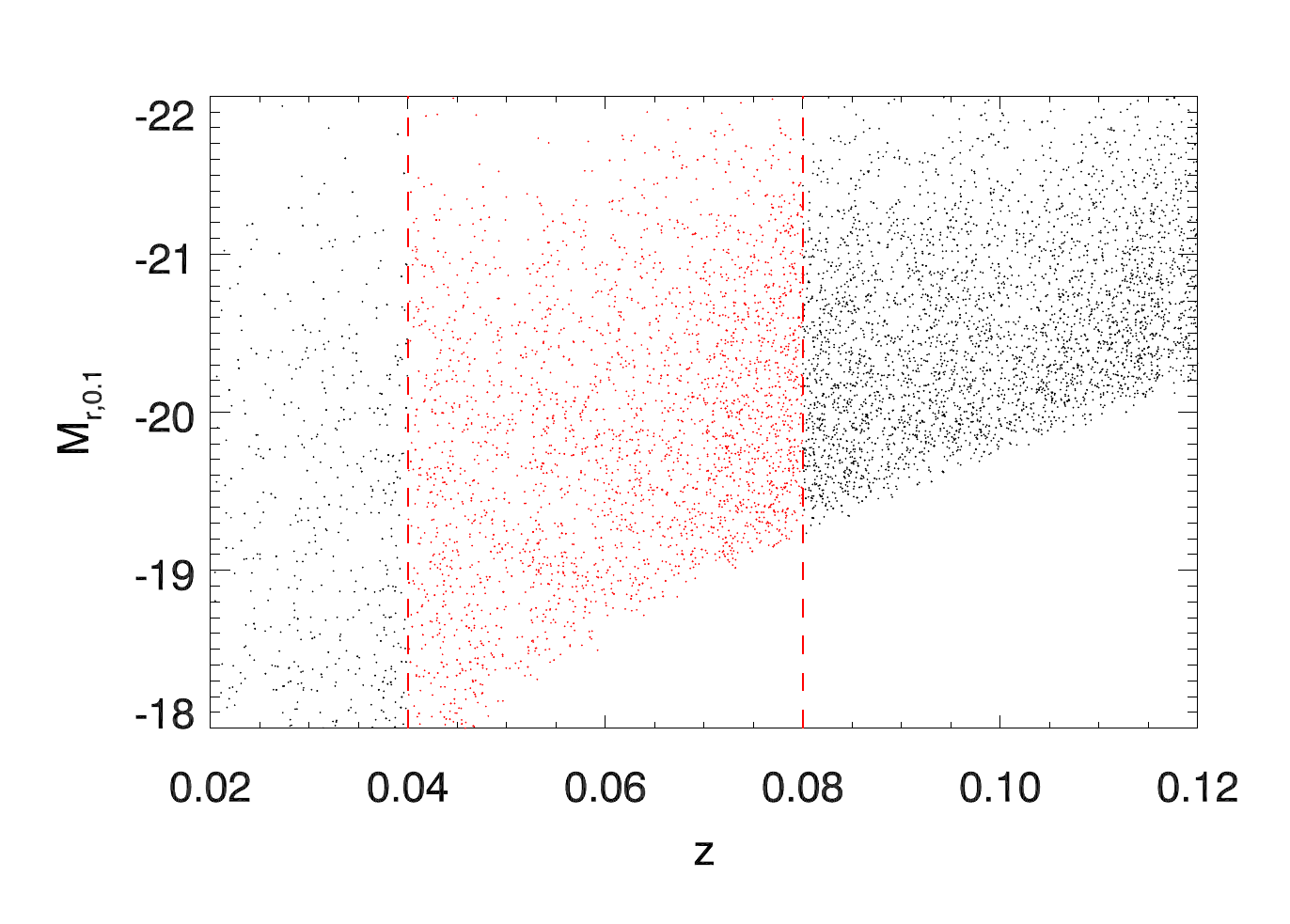}
\caption{A sub sample of the nearly face-on galaxies sample: $0.04<z<0.08$, galactic extinction at $u$ band less than 0.3 $mag$ and $r$ band less than 0.5 $mag$. Black dots show a random subsample of our main main sample, red dots show the subsample, this subsample contains 67499 galaxies, which is complete at $M_{r,0.1}<-19.2$.}
\label{fig.5}
\end{figure}

The mixture model provides a method of describing more complex probability distributions, by combining several probability distributions. Gaussian Mixture Model (GMM) is a parametric probability density function represented as a weighted sum of Gaussian component densities. GMM is among the most statistically mature methods for clustering \citep{Xu_1996, Dasgupta_1999, Verbeek_2003}.

When fitting models, it is possible to increase the likelihood by adding more free parameters, however, which may result in overfitting. To minimize the number of redundant parameters, we make use of Akaike information criterion (AIC) and Bayesian information criterion (BIC) for model selection. As shown in Figure 6, the AIC and BIC have the same trend, but BIC converges faster than AIC, so we prefer BIC as our criterion, and treat the minimum value of BIC as the best models for fitting. 

The outlier data in each diagram may affect the distribution of AIC and BIC and produce extra components that occupy very little fraction and spread distribution, it will generate one or two ellipses that have tiny centre and large radius, like a ``background" in each diagram. This is the intrinsic problem of BIC, and we can minimize this effect by removing outliers.

The results of GMM fitting are shown in Figure 6. For $(u-r)$ vs. $M_r$, the GMM exactly detects the blue cloud and red sequence. Between these two sequences, the green valley region contains two Gaussian components: one located in the faint end  of red sequence and extending to the faint end of blue cloud, and the other in the bright end of blue cloud stretching towards the bright end of red sequence. For the $(u-g)$ CMD, the red sequence is decomposed into two components, and the faint component in the green valley is bluer than the corresponding component in the $(u-r)$ CMD. 
Whether the number of components is N$=4$ or N$=5$, there are always two Gaussian components in the green valley region. In the 4th subgraph of Figure 6, the values of BIC are almost equal when N$=4$ and N$=5$, it suggests there is no much improvement for fitting when we chose five components, so we prefer the four components result as our finding.

\begin{figure*}
\includegraphics[width=7.0in]{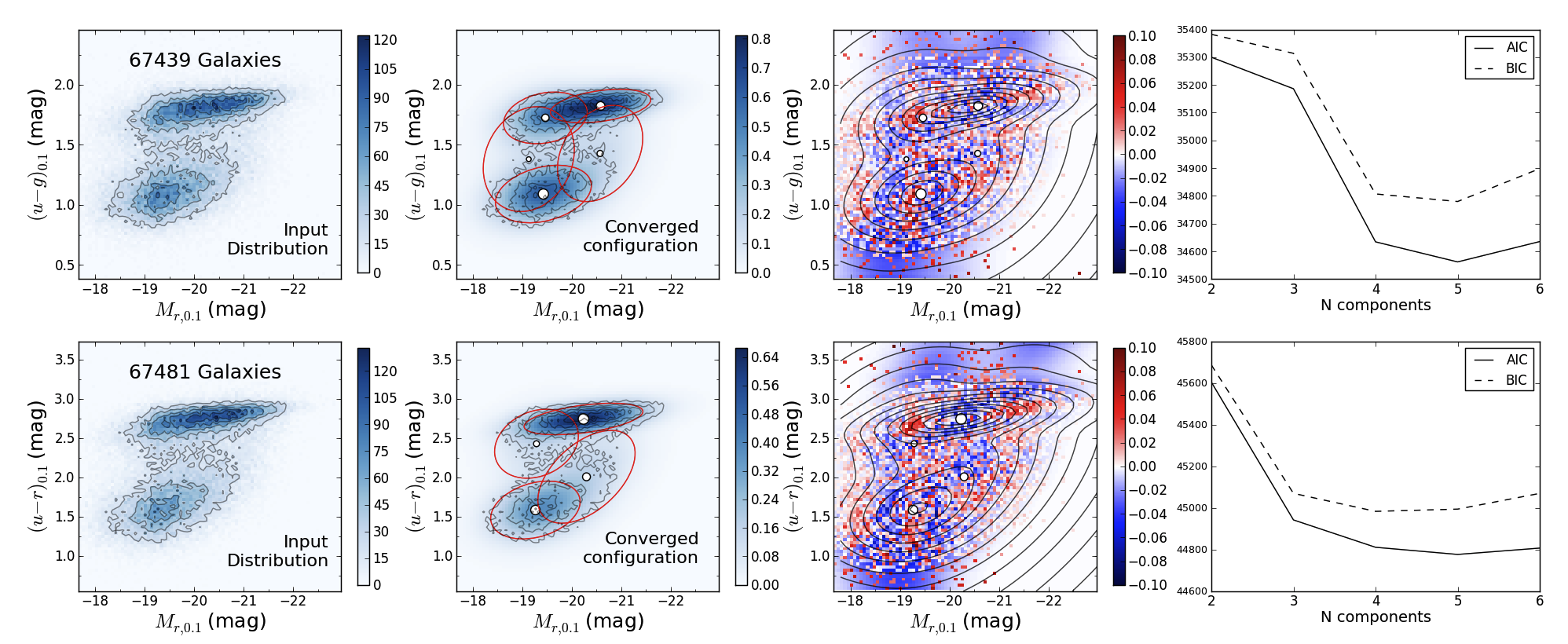}
\caption{The GMM fitting results of $(u-g)$ and $(u-r)$ CMDs, the first column plots the input CMDs; the second column shows the best-fitting Gaussian components, the red ellipse is the 1.5 $\sigma$ region of each Gaussian, the size of ellipse centre illustrates the fraction of each Gaussian component; the third column shows the residuals between input distribution and models; the final column plots the AIC and BIC, we prefer BIC to our criterion.}
\label{fig.6}
\end{figure*}

\begin{figure*}
\includegraphics[width=7.0in]{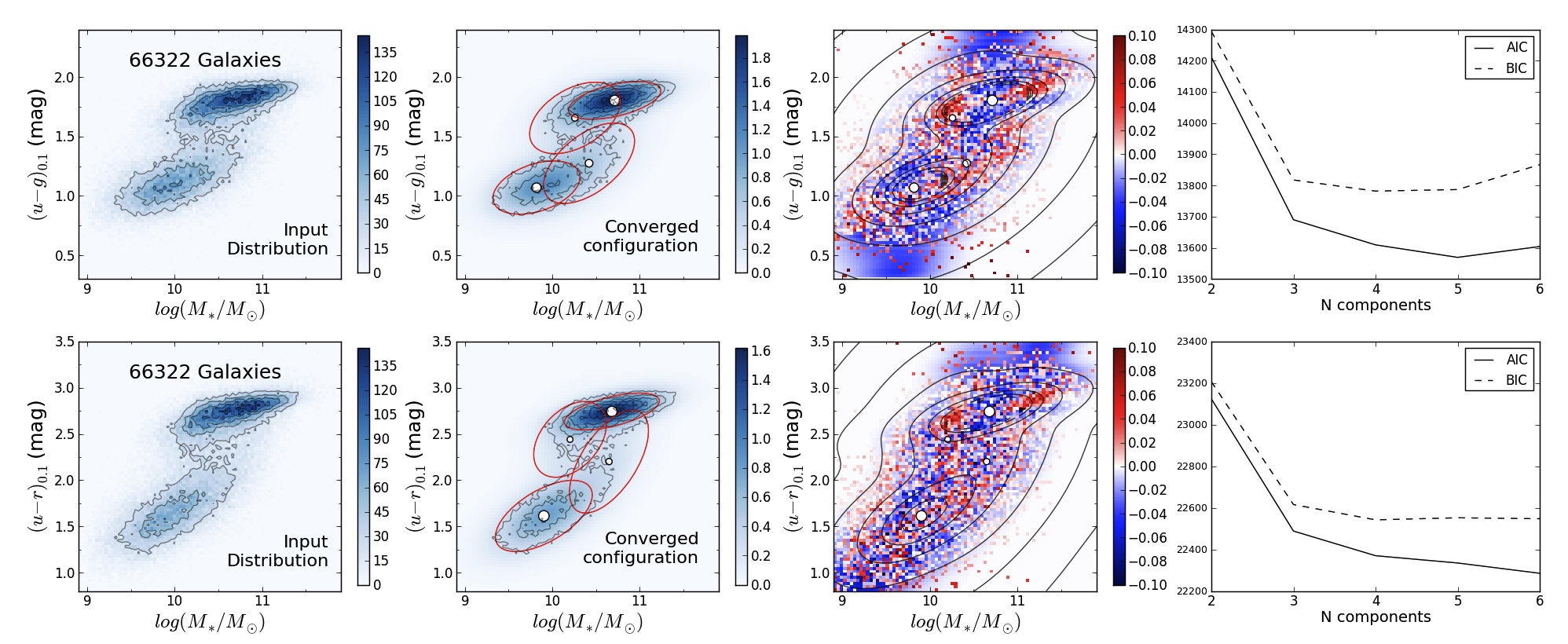}
\caption{The results of GMM fitting in color-stellar mass space.}
\label{fig.7}
\end{figure*}

Figure 7 shows the GMM results for color-mass space. The distribution of Gaussians is the same as that in Figure 6. As shown in Figure 7, the mean stellar mass of the bright component in the green valley region is close to the characteristic mass $M^*$ $(10^{10.5}M_\odot)$. 

Given that galaxies fainter than $M_{r,0.1}=-19.2$ $mag$ are not complete in this subsample, we check the results by only fitting the complete ($M_{r,0.1}<-19.2\ mag$) sample only, and find that the AIC and BIC converge at very large values. But if we set the same number of Gaussian components as in Figure 6, the ellipses' centre almost do not change while the size of ellipses are somewhat different. The completeness thus does not change the GMM fitting results.

Essentially, the work in section 3.1 is a 1-dimensional (1-D) fitting and GMM fitting is a 2-dimensional (2-D) fitting. The 1-D color fitting show that the color distribution of galaxies cannot be fitted by two Gaussians, which may suggest some potential components in the green valley region, and the GMM fitting successfully decomposes the color-magnitude distribution.

\begin{figure*}
\centering
\includegraphics[scale=0.53]{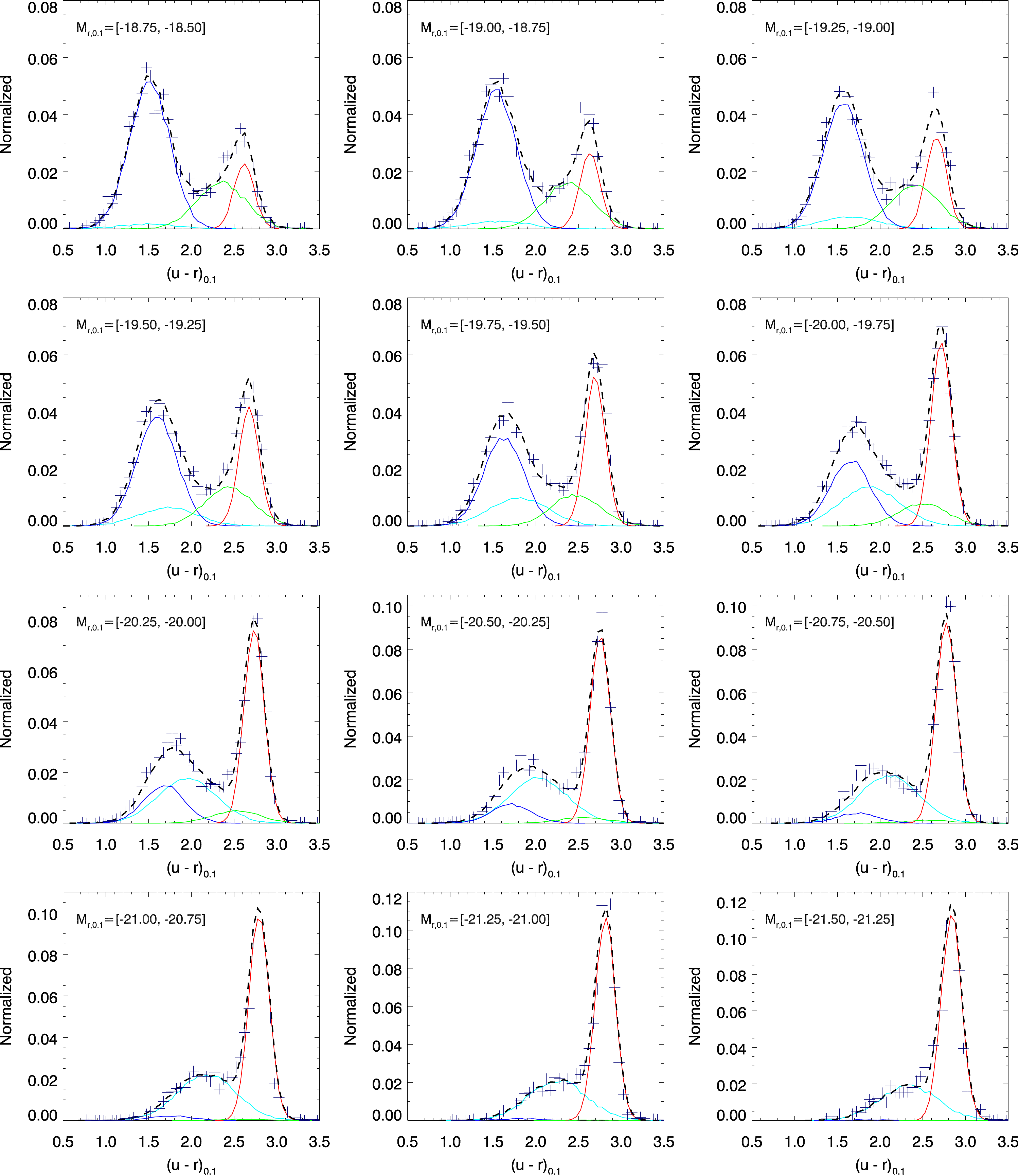}
\caption{ The color distribution of GMM result for galaxies in each magnitude bin. The color solid lines show the models generated by GMM fitting: 
the red and blue lines represent the red sequence and blue cloud, the green and cyan lines mark the faint and bright components of green valley respectively.
The black dashed line show the sum of models, and the plus signs mark the data.}
\label{fig.8}
\end{figure*}

As small residuals and well-distributed shown in Figure 6, GMM fit the data very well. We also show the GMM results in terms of 1-D color distribution in Figure 8, the color solid lines mark the models generated from GMM fitting: the red and blue lines represent the red sequence and blue cloud, the green and cyan lines mark the faint and bright components of green valley respectively. The sum of models (black dashed line) also fit the data (plus signs) very well too. Distinct from fitting in Section 3.1, the component number in Figure 8 is selected by BIC, and the model distribution (color solid line) in each magnitude bin is not a Gaussian, but a skew Gaussian because it is 1-D projection of an incomplete 2-D Gaussian. These are advantages of GMM fitting, which make its results reasonable.

\section{Discussion}

In this study we carried out Gaussian profile fittings to the CMD through two methods, i.e., the fitting to distributions in individual magnitude bins and to distributions in 2-D as a whole. The two Gaussian components in the CMD green valley are homologous to the excess in color distribution fitting, and explain why we cannot fit the color histogram well only two Gaussians. The green valley region is not dominated by a single Gaussian, but at least two Gaussian components. Galaxies in the green valley are composed of these two components plus the Gaussian tails of blue and red galaxies.

By fitting the $(u-r)$ color distribution in each magnitude bin, our results are different from those of \citet{Baldry_2004}, but agree with the $(NUV-r)$ results of \citet{Wyder_2007}.
As we select only nearly face-on galaxies which are not considered by \citet{Baldry_2004}, we attribute the difference to dust extinction, which seriously reddens the color of blue cloud galaxies and covers the excess in the green valley region. 

\citet{Kauffmann_2003a} showed that galaxies tend to divide into two distinct groups below and above a stellar mass of $3\times10^{10}M_\odot$. Galaxies below this mass limit tend to have younger stellar populations, while more massive galaxies tend to be older. We find that the fraction of blue cloud and red sequence are equal to each other around $M_{r,0.1}$$\sim$$-20.1\ mag$, corresponding to stellar mass about $M^*$ ($10^{10.5}M_{\sun}$). 
As shown in Figure 3, our result is consistent with their conclusion very well.

If we assume that the density of galaxies represents their evolution time scale in color-magnitude space, the GMM results would intuitively show us the evolutionary paths of different galaxies: the components in Figure 6 would represent galaxies in different evolving phases, and the inclination of ellipses' major axes may suggest the galaxy evolving direction in color-magnitude space.
The faint component in the green valley may be the early-quenching population, which quenched while galaxy are still small and grow mass along red sequence via ``dry" mergers \citep{Faber_2007}. 
In the faint end of the red sequence, these red low-luminosity galaxies tend to be in overdense regions \citep{Blanton_2006}, environment possibly plays a very important role in their evolution. 

If the two components in the green valley are dominated by two independent quenching processes, this scene would agree with \citet{Peng_2010} very well. According to the model of \citet{Peng_2010}, galaxies in the faint component are dominated by ``environment quenching", and galaxies in the bright one are dominated by ``mass quenching", 
which are associated with the quenching processes of satellite and central galaxies in group \citep{Peng_2012}.
Since these two effects are fully independent of each other, they may produce the two independent Gaussian components we found in the CMD.

As shown in Figure 7, the mean stellar mass of the bright component in the green valley region is about the characteristic mass $M^*$. The characteristic mass $M^*$ also corresponds to a dark halo mass $M_{shock}\sim10^{12}M_\odot$ based on the model of \citet{Dekel_2006}. In their model, galaxies with $M_{halo}\gtrsim10^{12}M_\odot$ will generate a steady shock in the gas accreting onto dark matter halo, the shock heats the gas and absolutely quenches star formation when AGNs begin to work. This process is strongly related to the mass of galaxy, which may dominate the evolution of bright galaxy component in green valley.

\citet{Wong_2012} selected a local post-starburst galaxies (PSGs) sample from SDSS, those PSGs occupy the low-mass end of the "green valley" below the transition mass within the colour-stellar mass diagram (the same position as the faint component of green valley in Figure 7). 
They proposed those PSGs represent a population of galaxies which is rapidly transitioning between the star-forming and the passively evolving phases.
\citet{Mendel_2013} select a sample of young passive galaxies from SDSS, which is identified based on the contribution of A-type stars to spectra and the relative lack of ongoing star formation. Most of these recently quenched galaxies have a stellar mass $> 10^{9.5}M_\odot$ and are predominantly early-type systems. \citet{McIntosh_2013} studied the recently quenched ellipticals (RQEs) with stellar mass $>10^{10}M_\odot$ and found a number of RQE properties are consistent with these galaxies being new remnants from a gaseous major merger. Their studies show that the low- and high- mass galaxies in green valley are very different, and suggest the green valley are dominated by different quenching processes, supporting our GMM fitting results.

In Figure 8, it is notable that the blue peak of color bimodality is dominated by the bright component of green valley (the cyan line) for galaxies $M_r < -20.25$ $mag$, which imply that the blue galaxy population with $M_*>10^{10.5}M_\odot$ is distinct from galaxies which traditionally thought as blue cloud.
This case is consistent with \citet{Schawinski_2014}, which proposed that the early- and late-type galaxies in green valley have two different evolutionary pathways. The evolving early-type galaxies generally have stellar mass $M_*>10^{10.5}M_\odot$, rapidly quenching star formation, moving out the blue cloud, into the green valley and to the red sequence as fast as stellar evolution allows \citep{Schawinski_2014}.

It is still unclear about the quenching mechanisms of galaxies, differentiating such mechanisms requires more evidences which are beyond the scope of this work. 

\section{summary}
In this paper, we have analyzed the color-magnitude distribution of a sample of nearly face-on galaxies, selected from SDSS DR7 main galaxy sample. We fit $(u-g)$ and $(u-r)$ color distribution in each magnitude bin and apply Gaussian Mixture Models to fit the CMDs and the results are as follows,

(1) The color distributions of galaxies cannot be fitted by two Gaussians, there is an obvious excess in the green valley region.

(2) With rising luminosity, red galaxies increasing while blue galaxies decrease, the fraction of blue cloud and red sequence cross at $M_{r,0.1}$$\sim$$-20.1\ mag$. At this magnitude, the mean stellar mass of galaxies is about the characteristic mass $M^*$. 

(3) By fitting the excess between blue cloud and red sequence, we find that the centre of the green-valley  for face-on galaxies in the local universe is: $(u-r)_{0.1}=-0.121M_{r,0.1}-0.061$.

(4) The GMMs of CMDs accurately illustrate the red sequence and blue cloud, and yields 2 Gaussian components for the green valley region, which might suggest two different evolutionary paths from blue cloud to red sequence.

\acknowledgments
The authors are very grateful to the anonymous referee for his/her thoughtful and instructive comments that significantly improved the content of this paper. 
This work is supported under the National Natural Science Foundation of China under grants (11273015, 11133001 and 11273060), the National Basic Research Program (973 program No. 2013CB834905), and Specialized  Research Fund for the Doctoral Program of Higher Education (20100091110009).

Funding for the SDSS and SDSS-II was provided by the Alfred P. Sloan Foundation, the Participating Institutions, the National Science Foundation, the U.S. Department of Energy, the National Aeronautics and Space Administration, the Japanese Monbukagakusho, the Max Planck Society, and the Higher Education Funding Council for England. The SDSS was managed by the Astrophysical Research Consortium for the Participating Institutions.

This publication made extensive use of the Python package AstroML and scikit-learn, which can be found at http://www.astroml.org/ and http://scikit-learn.org/.

\clearpage

\end{document}